\documentstyle[prb,preprint,eqsecnum,aps]{revtex}

\def\AmS{{\protect\the\textfont2
        A\kern-.1667em\lower.5ex\hbox{M}\kern-.125emS}}

\makeatletter
\tighten
\begin{document}
\draft
\title{Density expansion for the mobility in a quantum Lorentz model}

\author{K. I. Wysoki\'nski}
\address{Institute of Physics,\\
 M.Curie-Sk\l odowska University,\\
 20-031 Lublin, Poland\\
 and\\
 Materials Science Institute\\
 University of Oregon\\
 Eugene, OR 97403}
\author{Wansoo Park and D. Belitz}
\address{Department of Physics and Materials Science Institute,\\
 University of Oregon,\\
 Eugene, OR 97403}
\author{T. R. Kirkpatrick}
\address{Institute for Physical Science and Technology,\\
 University of Maryland,\\
 College Park, MD 20742}
\date{\today}
\maketitle
\begin{abstract}
We consider the mobility of electrons in an environment of static
hard-sphere scatterers, which provides a realistic description of
electrons in Helium gas. A systematic expansion in the scatterer
density is carried to second order relative to the Boltzmann
result, and the analytic contribution at this order is derived,
together with the known logarithmic term in the density expansion.
It is shown that existing experimental data are
consistent with the existence of the
logarithmic term in the density expansion, but more precise
experiments are needed in order to unambiguously detect it.
We show that our calculations provide the necessary
theoretical information for such an experiment, and give a
detailed discussion of a suitable parameter range.
\end{abstract}

\narrowtext

\section{Introduction}
\label{sec:I}

It has been known for almost thirty years that transport coefficients,
in contrast to thermodynamic quantities, do not possess a virial
expansion.\cite{DorfmanKirkpatrickSengers} This insight came
originally as a substantial surprise,\cite{Peierls}
and the physics behind it turned out to be very fundamental in nature.
Namely, it is related to the, previously
unexpected, existence of long-range dynamical correlations in
equilibrium fluids. As a specific example, let us consider the
diffusion of a tagged particle in a fluid. The classic method for
setting up a virial or density expansion for the diffusivity was
based on a generalized Boltzmann equation.\cite{Bogoliubov} In this
method one constructs a virial expansion for the collision operator
by taking into account collisions of the tagged particle with a
successively increasing number of scatterers which is, however,
finite at every order in the expansion. This is analogous to the
Ursell-Mayer cluster expansion for thermodynamic
quantities.\cite{Wannier} Alternatively, one can use the Green-Kubo or
time correlation function method.\cite{BoonYip} Both methods yield
identical formal results for the density expansion of the diffusivity
$D$ (or any other transport coefficient),
\begin{equation}
D/D_B = 1 + D_1 n + D_2 n^2 + O(n^3)\quad,
\label{eq:1.1}
\end{equation}
where $D_B$ is the Boltzmann result for the diffusivity, $n$ is the
dimensionless density of the fluid, and $D_1$, $D_2$, etc. are the
'virial coefficients'. The surprise consisted in the realization that
the coefficients in this expansion are infinite past a certain
order that depends on the dimensionality of the system.\cite{DorfmanCohen}
It was soon realized that these infinities signalize the presence of
a logarithmic density dependence in the expansion given by
Eq.\ (\ref{eq:1.1}).\cite{log} For instance, in a three-dimensional
(3-d) system, the coefficient $D_2$ is $n$-dependent, and goes for small
$n$ like $D_2 \sim \ln n + {\rm const}$, so that Eq.\ (\ref{eq:1.1}) must
be replaced by,
\begin{equation}
D/D_B = 1 + D_1 n + D_{2log} n^2\ln n + D_2 n^2 + o(n^2)\quad,
\label{eq:1.2}
\end{equation}
where $D_{2log}$ and $D_2$ are numbers, and $o(n^2)$ denotes terms that
vanish faster than $n^2$ for $n\rightarrow 0$. The functional form
of the higher coefficients is not known. The physical reason for the breakdown
of the density expansion, and the appearance of the logarithmic terms,
is a collective effect: Ring collisions, in which the scattered particle
collides with a scatterer, and then visits a number of other scatterers
before returning to the original one, turn out to lead to infinite
phase space volumes if a finite number of scatterers is considered, and
hence the distances between scatterers, and the times traveled between
them, are allowed to become arbitrarily large. These collision events
are, however, unphysical, because the inevitable presence of scatterers
that do not belong to the particular cluster under consideration does
not allow the effective distance traveled between successive scattering
events to be substantially longer than a mean-free path.
Properly taking into account this mean-free-path damping of the scattered
particle's trajectory leads to the mentioned logarithmic term. The latter
represents dynamic correlations in the fluid which extend over a distance
of a few mean-free paths (or a time equal to a few mean-free times),
which in the limit of small densities becomes
much larger than the effective particle size, or the range of the
intermolecular force. In contrast, static correlations in classical equilibrium
fluids extend only over the range of the intermolecular force,
which is the reason why the cluster expansion works for
thermodynamic quantities.

Due to its bearing on fundamental aspects of the Statistical Mechanics of
both equilibrium and non-equilibrium systems,
and its close connection to related effects, such
as long-time tails,\cite{DorfmanKirkpatrickSengers}
the breakdown of the density
expansion for transport coefficients has generated
substantial interest over almost thirty years. The existence of the
logarithmic term has been ascertained theoretically for a variety of
classical systems, both real fluids and model fluids,
and for a variety of transport coefficients.\cite{log} It has also
been seen in computer simulations of a 2-d Lorentz
gas,\cite{Bruin} where the logarithm appears at first order rather than
at second order due to the lower dimensionality. It has also
been established that the same effect occurs for quantum mechanical
scattered particles.\cite{LangerNeal}
However, its experimental verification has proven to be extraordinarily
difficult.\cite{LawSengers} The reasons for this are manifold. First,
the detection of any logarithmic term on an analytic background is
very difficult due to the slow variation of the logarithm. Second, for
3-d systems the effect appears only at second order in the density
expansion beyond the Boltzmann equation. Third,
the coefficients in Eq.\ (\ref{eq:1.2}), or in the equivalent expressions
for other transport coefficients, are not known for classical fluids
with realistic interaction potentials. Even for a hard-core model fluid
it has so far not been possible to calculate the coefficient $D_2$ of the
analytic term at second order. Since the detection of a logarithmic
term on an {\it unknown} background is a hopeless task, this essentially
precludes the use of 3-d classical fluids for a convincing observation.
On the other hand, 2-d classical systems are very hard to realize.
Furthermore, estimates of the values of the coefficients
in classical systems show that for those transport coefficients that
can be accurately measured, like e.g. the shear viscosity, the coefficient
of the analytic second order term is much larger than the one of the
logarithmic term. This is because the former
contains essentially excluded volume effects, which dominate
over the ring collision contributions.

These difficulties raise the question whether quantum systems are
possibly better suited for an experimental verification of the logarithmic
term than classical ones. 2-d quantum systems, although easy to realize,
are not suitable since transport in 2-d quantum systems is pathological
due to localization effects.\cite{LeeRama,R} This leaves 3-d quantum
systems. Here the effect also occurs only at second order, but otherwise
the situation is much more favorable than in the case of 3-d classical
systems, as we will see. A chief advantage is the fact that the
hard-core Lorentz
gas model, i.e. a tagged particle moving in an array of static hard-sphere
scatterers,\cite{Hauge} is a much better approximation for certain
quantum systems than for any classical ones. 3-d quantum systems are
therefore the most promising candidates for an experimental observation
of the nonanalyticity.

\par
A particularly promising system consists of electrons injected into Helium
gas of density $n$.\cite{Schwarz} The electron-Helium scattering process
is well known, and its characteristics are convenient from a theoretical
point of view. The scattering length, $a_{s}=0.63\AA$, is positive,
and for thermal electrons the energy dependence of the scattering cross
section is negligible. Since the electrons behave quantum mechanically,
the thermal wavelength, $\lambda = (2\pi^2\hbar^{2}\beta/m)^{1/2}$, with
$\beta=1/k_{B}T$ and $m$ the electron mass, provides an additional length
scale besides $a_{s}$ and the mean Helium atom separation $n^{-1/3}$.
The leading parameter in the density expansion is $na_{s}^{2}\lambda =
\lambda/4\pi l$,\cite{KirkpatrickDorfman} with $l=1/4\pi n a_{s}^{2}$ the
mean free path, and $a_{s}/\lambda$ serves as an additional small
parameter. The mass ratio $m_{He}/m\approx 10^{4}$
makes it a good approximation
to treat the Helium atoms as static scatterers. Finally, the low density
of the injected electrons allows one to neglect Coulomb interaction
effects between the electrons. An experiment \cite{Schwarz} which measures
the mobility of the electrons (which is related to the diffusion coefficient
by an Einstein relation) thus constitutes an almost ideal realization of a
$3-d$ quantum Lorentz model.

\par
The density expansion for the transport coefficients of the $3-d$
quantum Lorentz model has been considered
in Refs.\onlinecite{KirkpatrickDorfman,KirkpatrickBelitz}.
The experimentally relevant quantity is the mobility, $\mu$, at finite
temperature. The leading
terms in the expansion for $\mu$, analogous to Eq.\ (\ref{eq:1.2}), are,
\begin{mathletters}
\label{eqs:1.3}
\begin{equation}
\mu/\mu_{B} = 1 + \mu_{1}\,\chi + \mu_{2log}\chi^2\ln \chi +
                   \mu_{2}\chi^2 + O(\chi a_{s}/\lambda) + o(\chi^2)\quad,
\label{eq:1.3a}
\end{equation}
with
\begin{equation}
\mu_{1} = -\pi^{3/2}/6\quad,
\label{eq:1.3b}
\end{equation}
\begin{equation}
\mu_{2log} = (\pi^2-4)/32\quad.
\label{eq:1.3c}
\end{equation}
Here $\mu_B(T) = (4el/3)(2\pi m/\beta)^{-1/2}$ is the Boltzmann mobility,
and $\chi=\lambda/\pi l$ and $a_s/\lambda$ are small parameters.

\par
Adams et al.\cite{Adamsetal} have used Eqs.\ (\ref{eqs:1.3}) to
analyze experimental data obtained from time-of-flight measurements for
electrons in He and $\rm H_{2}$.
Their main objective was to refute the popular
misconception that $\mu_{1}=\mu_{2log}=0$,\cite{Rdensity} which
arose from an inappropriate application of localization ideas to the
low-density regime. Ref.\ \onlinecite{Adamsetal} concluded that the
existing experiments give very good agreement with the value of $\mu_{1}$
given in Eq.(\ref{eq:1.3b}). This success raised the question whether the same
system could be used to observe the logarithmic term. In the absence of
information about $\mu_{2}$ this would involve measuring the conductivity
over a gas density range that is sufficient to observe the logarithmic
dependence directly. This is clearly hopeless. However, if $\mu_{2}$
was known, then the logarithmic term would just provide a weakly density
dependent correction to it, and a sufficiently accurate experiment
{\it at fixed gas density} would be sufficient to probe the existence of
the logarithmic term. This background provided the motivation for a
calculation of $\mu_2$, the result of which has been reported
in a recent short communication as,\cite{Letter}
\begin{equation}
\mu_2 =  0.236\dots\quad.
\label{eq:1.3d}
\end{equation}
\end{mathletters}%

\par
The purpose of the present paper is to provide the technical details
of the calculation whose result was reported in Ref.\ \onlinecite{Letter}.
We will also give a more detailed discussion of how our result can
be used to design an experiment capable of observing the logarithmic
term. The paper is organised as follows. In Section \ref{sec:II} we first
define the model, and set up a diagrammatic perturbation theory for
the conductivity at zero temperature. We then perform the calculation
to second order in the scatterer density, and convert the result into
an expression for the experimentally relevant temperature dependent
mobility. Section \ref{sec:III} contains a detailed discussion of the
relevance of our results for a proposed experiment to detect the
logarithmic term in the density expansion. Some technical details are
relegated to two appendices.

\section{DENSITY EXPANSION FOR THE MOBILITY}
\label{sec:II}

\subsection{The Model}
\label{subsec:II.A}

We consider a model of noninteracting electrons moving in
three dimensional space, and scattering off
static, uncorrelated, and randomly located impurities
with density $n$. The electron-impurity
interaction can be characterised by the scattering cross section
$\sigma_{s} = 4\pi\,a^{2}_{s}$, where $a_{s}$ is the
scattering length.

The Hamiltonian for this system is,
\begin{equation}
 H = \sum_{{\bf k},\sigma} \left(\epsilon_{\bf k}
          - \mu \right) a^{\dagger}_{{\bf k},\sigma}a^{\,}_{{\bf k},\sigma}
     + \sum_{{\bf k},{\bf q},\sigma} \, V({\bf q}) \,
           a^{\dagger}_{{\bf k},\sigma} a_{{\bf k}-{\bf q},\sigma}\quad,
\label{eq:2.1}
\end{equation}
where $a^{\dagger}_{{\bf k},\sigma}$, $a_{{\bf k},\sigma}$ denote
creation and
annihilation operators for the electron with wavenumber ${\bf k}$
and spin $\sigma$, $\mu$ is the chemical potential, $V({\bf q})$ is
the Fourier transform of the electron-impurity scattering
potential, and $\epsilon_{\bf k} = {\bf k}^2/2m$ is with $m$
the electron mass. In this section we use units such that $\hbar=1$.

We will use the standard Edwards diagram technique.\cite{AGD}
Accordingly, we consider retarded (R) and advanced (A)
zero-temperature Green's functions,
\begin{mathletters}
\label{eqs:2.2}
\begin{equation}
{\cal G}^{R,A}_{{\bf k},{\bf p}} (\omega) = \left\langle {\bf k} \left\vert
   \frac{1}{\omega - H \pm i0} \right\vert{\bf p}\right\rangle
\label{eq:2.2a}
\end{equation}
and their impurity averaged counterparts,
\begin{equation}
\langle {\cal G}^{R,A}_{\bf k,p}(\omega)\rangle
   = \delta_{\bf k,p}\ \frac{1}{\omega + \epsilon_F - \epsilon_{\bf k}
                                       + \Sigma^{R,A}_{\bf k}(\omega)}\quad,
\label{eq:2.2b}
\end{equation}
\end{mathletters}%
where $\epsilon_F = \mu (T=0)$ is the Fermi energy, and
$\Sigma^{R,A}_{\bf k}(\omega)$ is the self energy. $\omega$ measures
the energy distance from the Fermi surface. Ultimately we will take the
limit $\omega\rightarrow 0$ to obtain the static mobility or conductivity.

We will restrict our considerations to pure s-wave scattering, which is
equivalent to the assumption of point-like scatterers. This approximation
will substantially simplify all calculations. Its justification and
limitations will be discussed in Sec.\ \ref{sec:III}. For pure s-wave
scattering the self-energy reads to linear order in the impurity
density,
\begin{equation}
\Sigma_{\bf k}^{R,A} (\omega) = \Delta \pm i/2\tau + O(n^2)\quad,
\label{eq:2.3}
\end{equation}
with $\tau = lm/k_F$ the mean-free time. The real part of the
self energy, $\Delta$, is to this order independent of the
wavenumber ${\bf k}$. Even though it is formally infinite for pure
s-wave scattering, it therefore strictly
renormalizes the chemical potential or the zero of energy, and can
be neglected. Although we will want to effectively expand the
full Green's function to second order in the impurity density, we
will find it convenient to use the self energy as given by
Eq.\ (\ref{eq:2.3}) and to include higher order contributions to
the self energy explicitly at a later stage. We therefore define,
for later reference, an auxiliary zero frequency Green's function,
\begin{mathletters}
\label{eqs:2.4}
\begin{equation}
G^{R,A}_{\bf k} = \frac{1}{\epsilon_F - \epsilon_{\bf k} \pm i/2\tau}\quad.
\label{eq:2.4a}
\end{equation}
We will also need the free electron Green's function,
\begin{equation}
G^{(0)R,A}_{\bf k} = \frac{1}{\epsilon_F - \epsilon_{\bf k} \pm i0}\quad.
\label{eq:2.4b}
\end{equation}
\end{mathletters}%

\subsection{Diagrammatic Expansion of the Conductivity}
\label{subsec:II.B}

It is convenient to calculate the conductivity, $\sigma$,
of degenerate electrons at $T=0$,
and then to convert to the experimentally relevant finite-$T$ mobility by
means of an Einstein relation and a Kubo-Greenwood formula.
A general scheme for a diagrammatic calculation of the conductivity
within the context of the model defined in Sec.\ \ref{subsec:II.A}
above has been given in Ref.\ \onlinecite{KirkpatrickDorfman}.
The result was a density expansion for the conductivity of the form,
\begin{equation}
\sigma/\sigma_B = 1 + \sigma_1\ \frac{1}{2k_F l} +
          \sigma_{2log}\left(\frac{1}{2k_F l}\right)^2
                                   \ln\left(\frac{1}{2k_F l}\right) +
          \sigma_2\left(\frac{1}{2k_F l}\right)^2 +
          O(a_s/l) + o\left(1/(k_F l)^2\right)\quad,
\label{eq:add}
\end{equation}
with $\sigma_B$ the Boltzmann conductivity, $k_F$ the Fermi wavenumber,
and $a_s k_F$ considered small. There is no need to repeat the
description of the general scheme here.  Suffice it to say that
the real part of the dynamical conductivity of noninteracting electrons
at zero temperature can be expressed in
terms of the Green's functions defined in Eq.\ (\ref{eqs:2.2})
by means of the Kubo--Greenwood formula,
\begin{equation}
{\rm Re} \sigma(\omega) = \frac{e^{2}}{\pi m^{2}} \, {\rm Re} \sum_{\bf k,p}\
v({\bf k})\ \left\langle {\cal G}^{R}_{\bf k,p}(\omega)\,
{\cal G}^{A}_{\bf p,k}(\omega =0) -{\cal G}^{R}_{\bf k,p}(\omega)\,
{\cal G}^{R}_{\bf p,k}(\omega =0) \right\rangle\ v({\bf p})\quad,
\label{eq:2.5}
\end{equation}
where $v({\bf k}) = {\bf k\cdot q}/q$, with ${\bf q}$ an arbitrary
fixed vector, is the current
vertex. The brackets $\langle\cdots\rangle$ denote the averaging
over the random positions of the impurities. We will perform this average
by means of standard diagrammatic methods.\cite{AGD} We note that
this is not the only possible way to calculate the conductivity.
One could, for instance, use quantum kinetic theory instead. However,
it is known from calculations of the coefficient $\sigma_1$ in
Eq.\ (\ref{eq:add}) that the diagrammatic method
\cite{KirkpatrickDorfman,KirkpatrickBelitz} leads to much
simpler calculations than quantum
kinetic theory.\cite{KirkpatrickDorfmanI} Technically, our calculation is a
systematic extension of the work of Refs.\ \onlinecite{KirkpatrickDorfman}
and \onlinecite{KirkpatrickBelitz}. These authors had already identified
all diagrams that contribute to $\sigma_1$ and $\sigma_{2log}$, and
calculated both coefficients. For our purposes we will now have to calculate
the previously considered diagrams to $O(n^2)$ instead of $O(n^2\ln n)$,
and to identify and evaluate all diagrams that contribute to $\sigma_2$,
but neither to $\sigma_1$ nor to $\sigma_{2log}$.

Let us start by introducing symbols and abbreviations for the
diagrammatic elements of our perturbation theory. The averaged Green's
function, Eq.\ (\ref{eq:2.4a}), we denote by a directed full line, with
an arrow pointing right and left for the retarded and advanced Green's
function, respectively. We further denote the difference between the
Green's function given by  Eq.\ (\ref{eq:2.4a}) and the free electron
Green's function,  Eq.\ (\ref{eq:2.4b}), by a directed line which
carries a triangle. The impurity potential is denoted by a dashed
line, and the impurity density by a cross; each cross with a dashed
line running through it corresponds to a factor $u=1/2\pi N_F\tau$, with
$N_F=k_F m/2\pi^2$ the free electron density of states per spin at the
Fermi level. The current vertex $v({\bf k})$ is represented by a
triangle. All of these diagrammatic elements can be seen in
Fig.\ \ref{fig:1}. We also define the quantities
$\epsilon \equiv 2m\epsilon_F$, and $\gamma \equiv m/\tau$.

In terms of these quantities, the simplest diagrammatic contribution
to Eq.\ (\ref{eq:2.5}) in the limit of zero frequency is the simple
bubble shown in Fig.\ 1(a), and the corresponding diagram
with the direction of the lower line reversed. The analytic expression
for this contribution to the static conductivity, which we denote by
$\sigma_{(1a)}$, reads,
\begin{mathletters}
\label{eqs:2.6}
\begin{equation}
\sigma_{(1a)} = \frac{e^{2}} {\pi m^{2}} \, {\rm Re} \, \sum_{\bf k}\
[v({\bf k})]^2\ \left[G^{R}_{\bf k}\,G^{A}_{\bf k} -
G^{R}_{\bf k}\,G^{R}_{\bf k}\right]\quad.
\label{eq:2.6a}
\end{equation}
Performing the integral is simple, and yields,
\begin{equation}
\sigma_{(1a)} = \sigma_{B} \left[1 + \frac{3}{2} \left(\frac{\gamma}
 {2\epsilon}\right)^{2} +
 O\left(\left(\frac{\gamma} {\epsilon}\right)^{4}\right)\right]\quad.
\label{eq:2.6b}
\end{equation}
Diagram 1(a) thus contributes $3/2$ to the coefficient $\sigma_2$, a
result which we record in Table \ref{tab:1}.
For later reference we also give explicitly the Boltzmann result for
the conductivity,
\begin{equation}
\sigma_{B} = \frac{e^{2}\epsilon^{3/2}}
  {3\pi^{2}\gamma} = \frac{e^{2}n_{e}\tau} {m} \quad,
\label{eq:2.6c}
\end{equation}
\end{mathletters}%
with the free electron density
$n_{e} = \epsilon^{3/2}/3\pi^{2}$, and $e$ the electron charge.
Notice that $\sigma_B$ is a function of
$\epsilon$, this will be of importance later. We also note that
the simple bubble is the only diagram for which we have to consider
the ${\cal G}^R {\cal G}^R$ contribution to Eq.\ (\ref{eq:2.5}).
All other diagrammatic contributions to this term turn out to
be of higher than second order in the impurity density.

Figure \ref{fig:1} also shows two diagrams that have to be considered
due to our choice of the basic Green's function, Eq.\ (\ref{eq:2.4a}).
If we had calculated the self energy in the basic Green's function
to second order in the impurity density these diagrams would not
appear. Note the appearance of the 'triangulated' Green's function,
which is necessary to avoid double counting. The evaluation of these
diagrams is also straightforward, and the results are given in Table
\ref{tab:1}.

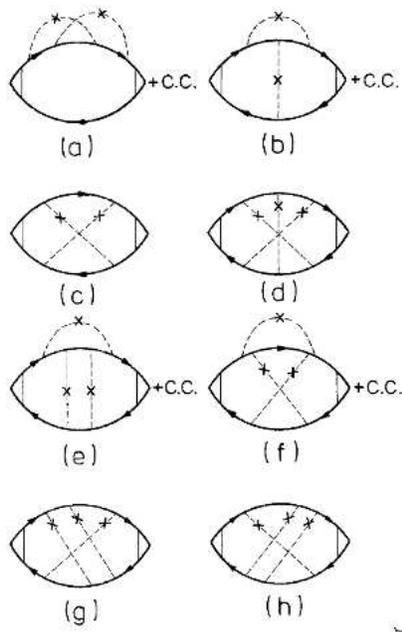
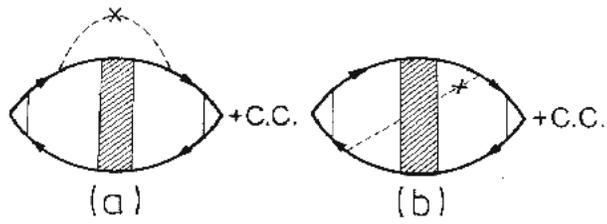
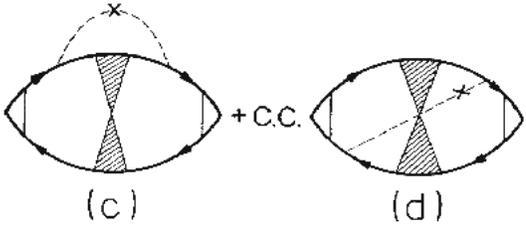
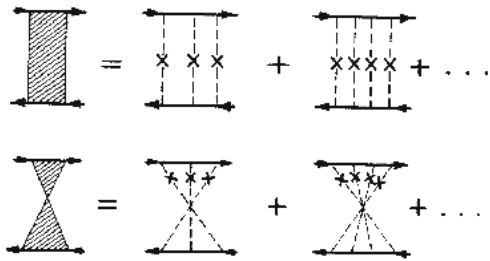
\begin{figure}
\caption{The diagrams that were considered previously in
 Refs.\ \protect\onlinecite{KirkpatrickDorfman,KirkpatrickBelitz}.}
\label{fig:2}
\end{figure}

We now turn to the diagrams that were discussed previously in Refs.\
\onlinecite{KirkpatrickDorfman,KirkpatrickBelitz}, and were calculated
there to $O(n^2 \ln n)$. They are shown
again in Fig.\ \ref{fig:2}. We find that all of these diagrams
contribute to the analytic term at second order as well. The
diagram shown in Fig.\ 2(a) is of special interest, since
for pure s-wave scattering it is found to be ultraviolet divergent.
We therefore introduce an ultraviolet cutoff $Q\sim 1/a_s$, and
discuss this diagram in some detail. We stress that this divergence
is due to an unphysical treatment of the short-range part of the
electron-impurity interaction and often occurs when one uses an
s-wave scattering approximation. It should not
be confused with the physically more interesting logarithmic
singularity in the density expansion of the transport coefficients
which is due to long-range collective effects.

The analytic expression corresponding to Fig.\ 2(a) is,
\begin{eqnarray}
\sigma_{(2a)} = 2u^{2} \sum_{\bf k,p,q} [v({\bf k})]^2\
 {\rm Re}\ \left(G^{R}_{\bf k}\,G^{R}_{\bf k-p}\,
 G^{R}_{\bf k-p-q}\,G^{R}_{\bf k-q}\,G^{R}_{\bf k}\,
 G^{A}_{\bf k}\right)\nonumber\\
 = \sigma_{B} \left(\frac{\gamma}{2\epsilon}\right)
      \frac{1}{2\pi^5\epsilon^{3/2}} \,
  {\rm Re} \int^{\infty}_{0} dq \, q^{2}\ J^{++}(q)\ \biggl[(\epsilon -
     i\gamma)\ [J^{++}(q) - J^{+-}(q)]\nonumber\\
          - (\epsilon + i\gamma)\ \gamma\ {d\over d\gamma}\,
    J^{++}(q)\biggr]\quad.
\label{eq:2.7}
\end{eqnarray}
To obtain the second equality we have repeatedly used the identity
\begin{equation}
G^{R}_{\bf k}\ G^{A}_{\bf k} = \frac{im}{\gamma}\
 \left(G^{R}_{\bf k} - G^{A}_{\bf k}\right)\quad,
\label{eq:2.8}
\end{equation}
and defined two functions,
\begin{mathletters}
\label{eqs:2.9}
\begin{equation}
J^{+-}(q) = \int d{\bf k} \frac{1}{\epsilon - {\bf k}^2 + i\gamma} ~~
 \frac{1} {\epsilon - ({\bf k} - {\bf q})^{2} - i\gamma}\quad,
\label{eq:2.9a}
\end{equation}
and
\begin{equation}
J^{++}(q) = \int d{\bf k} \frac{1}{\epsilon - {\bf k}^2 + i\gamma} ~~
 \frac{1} {\epsilon - ({\bf k} - {\bf q})^{2} + i\gamma}\quad.
\label{eq:2.9b}
\end{equation}
\end{mathletters}%
The same functions had been defined in
Refs.\ \onlinecite{KirkpatrickDorfman,KirkpatrickBelitz}, and all diagrams
can be expressed in terms of them. For our present purposes we need
a more accurate evaluation of these functions than the one that was
given before. It turns out that both integrals can be done exactly
in closed form, a task which we relegate to Appendix \ref{app:A}.
Using the result in Eq.\ (\ref{eq:2.7}), a calculation sketched in
Appendix \ref{app:B} leads to,
\begin{equation}
 \sigma_{(2a)} = \sigma_{B} \left[ - \pi\left(\frac{\gamma}
 {2\epsilon}\right) + \left(\frac{\gamma}
 {2\epsilon}\right)^{2}
 \left(6 \ln(Q/\sqrt{\epsilon}) + 4 - 12 \ln 2\right) +
 O \left(\left(\frac{\gamma}{2\epsilon}\right)^{3}\right)\right]\quad.
\label{eq:2.10}
\end{equation}
The $\ln Q$-contribution stems from the real part of the self-energy
contribution to the upper electron line in diagram 2(a). It thus
constitutes just a shift (albeit an infinite one in the case of a
pointlike potential) of the chemical potential, like the constant
$\Delta$ in Eq.\ (\ref{eq:2.3}) which we neglected earlier. We therefore
expect the $\ln Q$-term to disappear upon considering the experimentally
relevant mobility instead of the conductivity. We will find this
expectation to be borne out later, cf. Sec.\ \ref{subsec:II.C}.
The remaining diagrams
in Fig.\ \ref{fig:2} can be calculated along the same lines, and
the results are given in Table \ref{tab:1}.

The diagrams shown in Figs.\ 2(d)-(h) allow for generalizations which
all contribute to $\sigma_2$. These generalizations are obtained by
replacing the 'ladder' and 'crossed-ladder' elements in these
diagrams by the respective infinite resummations. The resulting
diagrams are shown in Fig.\ \ref{fig:3}. Note that the ladder,
or 'diffuson', and crossed-ladder, or 'Cooperon', resummations
in Fig.\ \ref{fig:3} start with three rungs each to avoid double
counting of the diagrams of Fig.\ \ref{fig:2}. The infinite
resummations are again easily expressed in terms of integrals over
the functions $J^{++}$ and $J^{+-}$, and the results are listed
in Table\ \ref{tab:1}. The reason why these diagrams with increasing
numbers of impurity lines, and hence increasing numbers of factors
$\gamma\sim n$, all contribute to the same order lies in
the fact that the diffusion pole contained in the ladder and
crossed-ladder resummations leads, with increasing order,
to increasingly singular infrared behavior of the integrand, which is
cut off only by $\gamma$. The cancellation of these two effects
leads to all of these diagrams being of the same order in $\gamma$
or $n$. By the same argument it follows that these are the {\it only}
infinite resummations (given our definition of skeleton diagrams by
means of Eq.\ (\ref{eq:2.4a})) that contribute to the desired order.
In particular, diagrams that contain more than one diffusion pole
do not contribute.

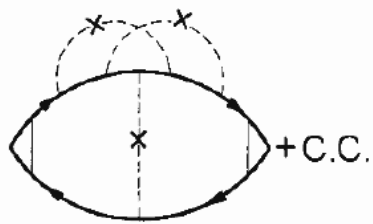
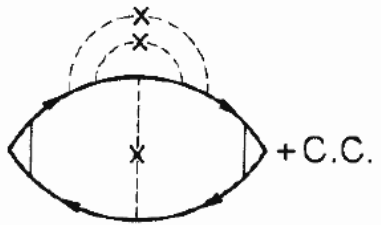
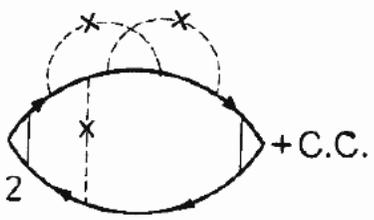
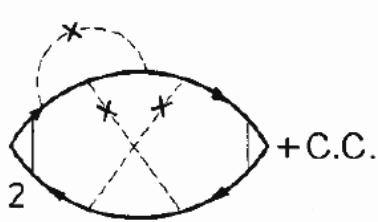
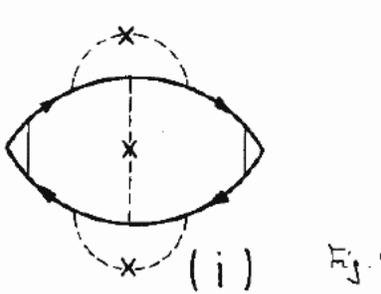
\begin{figure}
\caption{Skeleton diagrams with three impurity lines}
\label{fig:4}
\end{figure}

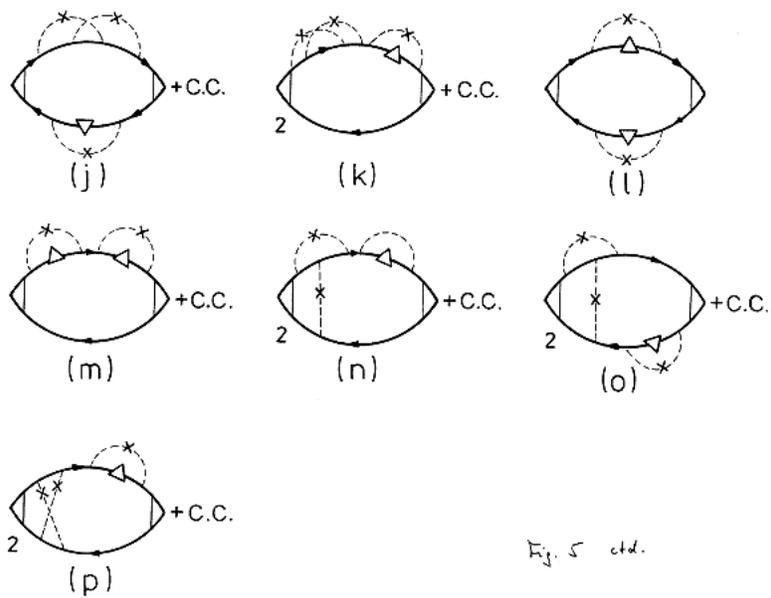
\begin{figure}
\caption{Skeleton diagrams with four impurity lines}
\label{fig:5}
\end{figure}

We now turn to other skeleton diagram contributions. All relevant
diagrams with three impurity lines (as far as they were not included
in Fig.\ \ref{fig:2}) are shown in Fig.\ \ref{fig:4}, and all those
with four impurity lines are shown in Fig.\ \ref{fig:5}. The evaluation
of these diagrams offers no particular difficulties, except that some
care has to be exercised since the vector nature of the current vertex
leads to some nontrivial angular integrations. Of course, all diagrams
that contain the same self-energy piece as diagram 2(a) also contain
the $\ln Q$-contribution that is characteristic for this diagrammatic
element. The results are again given in Table \ref{tab:1}.

Table \ref{tab:1} lists the contributions of all diagrams to the
coefficient $\sigma_2$ in Eq.\ (\ref{eq:add}). In some cases we have
found it convenient to combine some diagrams before evaluation of the
integrals, and this is indicated in the table. The coefficients
$\sigma_1$ and $\sigma_{2log}$ have been calculated in
Ref. \onlinecite{KirkpatrickBelitz} with the result,
\begin{equation}
\sigma_1 = -4\pi/3\qquad,\qquad\sigma_{2log} = (\pi^2-4)/2\quad.
\label{eq:addadd}
\end{equation}
Table \ref{tab:1} contains three integrals
which we could not reduce to tabulated ones, viz.
\begin{mathletters}
\begin{equation}
I_{1} = \int_{0}^{1} {dx \over x}\ \left[\ln \left({1-x \over 1+x}\right)
                                         \right]^{2} = 4.207\dots\quad,
\label{eq:2.11a}
\end{equation}
\begin{equation}
I_{2} = \int_{0}^{1} dx\ x\ \left[\ln \left({1-x \over 1+x}\right)
                                            \right]^{2} = 2.772\dots\quad,
\label{eq:2.11b}
\end{equation}
\begin{equation}
I_{3} = {4 \over \pi} \int_{0}^{\infty} {dx \over x^{2}}\
          (\arctan x)^{4}\ \left[1 - {1 \over x} \arctan x\right]^{-1}
                                                     = 7.716\dots\quad.
\label{eq:2.11c}
\end{equation}
\end{mathletters}%

Summing all the contributions listed in Table \ref{tab:1} we obtain
\begin{equation}
\sigma_{2} = 4 \ln(Q/k_{F}) + {55 \over 36}\pi^{2} - 14\ln 2 +
    7 - I_{1} + I_{2} - I_{3} = 4\ln(Q/k_F) + 3.22\dots\quad.
\label{eq:2.12}
\end{equation}

This concludes our calculation of the zero-temperature conductivity.
In order to compare with, and make explicit predictions for, experiments
it is desirable to convert this result into the corresponding one for
the mobility at nonzero temperature. In order to do this, we will
need the density of states to second order in the impurity density
as well as the conductivity. The density of states is easily obtained
from the Green's function via the relation,
\begin{equation}
N(\epsilon) = -\frac{1}{\pi}  \sum_{\bf k,p} {\rm Im} \
    \left\langle{\cal G}_{\bf k,p}^R (\omega=0)\right\rangle\quad,
\label{eq:2.13}
\end{equation}
and to second order it is sufficient to
consider the diagrams shown in Fig.\ \ref{fig:6}. The calculation
is easy, and we obtain,
\begin{equation}
 N(\epsilon) = N^{(0)}
  \left[1 + \left(\frac{\gamma}{2\epsilon}\right)^{2}
  \left(-\frac{1}{2} + 2\ \ln (Q/2\sqrt{\epsilon}) -
  2\ \ln 2\right) + O (\gamma^{3})\right]\quad.
\label{eq:2.14}
\end{equation}
Here $N^{(0)} = 2 N_F = mk_F/\pi^2 = m\sqrt{\epsilon}/\pi^2$
is the free electron density of states.

\subsection{The Electron Mobility}
\label{subsec:II.C}

In a time-of-flight experiment like the one described in
Ref.\ \onlinecite{Schwarz} the measured observable is the electron
mobility, which is given by $\mu(T)=\sigma(T)/en(T)$, where $\sigma(T)$
is the temperature dependent conductivity, and $n(T)$ is the electron
particle number density. $\sigma(T)$ can be obtained from Eq.(\ref{eq:add})
by means of the Kubo-Greenwood formula,\cite{Greenwood} and $n(T)$ from
the density of states, Eq.\ (\ref{eq:2.14}). We thus have,
\begin{equation}
\mu(T) = \int_{0}^{\infty} d\epsilon \biggl({-\partial f \over
\partial \epsilon}\biggr) \sigma(\epsilon)\bigg/e\int_{0}^{\infty}
d\epsilon f(\epsilon) N(\epsilon)\quad,
\label{eq:2.15}
\end{equation}
with $\sigma(\epsilon)$ and $N(\epsilon)$ from Eqs.\ (\ref{eq:add}),
(\ref{eq:addadd}), (\ref{eq:2.12}).
and (\ref{eq:2.14}), respectively, and $f(\epsilon)$ the Fermi function.
In writing all of these quantities as functions of
$\epsilon = k_{F}^{2}$ one has to keep in mind that the
electronic mean free path $l$ is the energy independent parameter
which has to be kept fixed. We are interested in the limit
of small electron density, where the Fermi function can be replaced by a
Boltzmann distribution,
$f(\epsilon)\approx \exp({\beta \mu})\ exp({-\beta \epsilon})$.
Doing the integrals we obtain Eq.\ (\ref{eq:1.3a}) for the temperature
dependent mobility with $\mu_1$ and $\mu_{2log}$ as given by
Eqs.\ (\ref{eq:1.3b},\ref{eq:1.3c}), and
\begin{equation}
\mu_{2} = {1 \over 16}\left[
{\pi^{2} \over 36}(55+9C) - C + 8 - 10\ln 2
-I_{1} + I_{2} - I_{3}\right] - 2\mu_{2log}\ln 2 = 0.236\dots\quad,
\label{eq:2.16}
\end{equation}
where $C$ is Euler's constant. As expected (see the discussion after
Eq.\ (\ref{eq:2.10})) the mobility is independent of the cutoff $Q$.

\section{Discussion}
\label{sec:III}

Since the mobility is directly measured in a time-of-flight experiment
of the type reported in Ref.\ \onlinecite{Schwarz}, the density
expansion for the mobility as given by Eqs.\ (\ref{eqs:1.3}), (\ref{eq:2.16})
can be
directly compared with experiment. We split the discussion of the
experimental relevance of our result into three separate questions:
(1) How accurately does our quantum Lorentz model describe electrons
injected into Helium gas? (2) How do our results compare with
existing data? (3) What kind of experimental effort would be necessary
to unambiguously detect the logarithmic term in the density expansion?

The first question has to be divided into idealizations that are
inherent in the model, and effects of additional approximations we
have made. Let us start out with the latter. As mentioned in the
Introduction, there are three independent length scales in the
model, viz. the mean scatterer separation $n^{-1/3}$, the mean-free
path $l$, and the thermal wavelength $\lambda$. Consequently, one can
form four different dimensionless densities, viz. $n\lambda^3$,
$n\lambda^2 a_s$, $n\lambda a_s^2$, and $na_s^3$. The first two do
not appear in the Lorentz model.\cite{KirkpatrickDorfmanI} Of the
other two, the first one is essentially the ratio of the thermal
wavelength to the mean-free path, while the second one describes an
excluded volume effect that is also present in classical systems.
We have kept only the leading (for $\lambda/a_s >> 1$) one of these two
parameters, $n\lambda a_s^2$. That is, we have neglected terms of
relative order $a_s/\lambda$. At Helium temperature, and with the
electron-Helium scattering length ($a_s = 0.63\AA$), the value of
this small parameter is $a_s/\lambda \approx 10^{-3}$.
The prefactor can be estimated from classical Enskog theory,
or from the exactly known result for the classical Lorentz model,
\cite{vanLeeuwenWeyland}, and turns out to be of order unity.
On the other
hand, a typical value for the expansion parameter $\chi$ in
Eq.\ (\ref{eq:1.3a}) is $\chi \approx 0.1$.\cite{Schwarz} In our
expansion, excluded volume corrections to the
term linear in $\chi$ will therefore roughly be of the same order
as terms $\sim \chi^3$. We can thus safely neglect excluded volume
effects, unless $\chi$ is chosen to be too small, see below.
It is also worthwhile to point out that the excluded
volume terms have a different temperature dependence than the
leading terms that were kept in Eq.\ (\ref{eq:1.3a}). In principle
it would therefore be possible to separate these contributions
experimentally, although this may be hard to do in practice.

Another approximation has been our restriction to pure
s-wave scattering. This amounts to neglecting corrections of
$O(k_F/Q)\sim O(k_F a_s)$ in Eq.\ (\ref{eq:add}), which translates
into corrections of $O(a_s/\lambda)$ in Eq.\ (\ref{eq:1.3a}).
Non-s-wave scattering effects and excluded volume effects are therefore
comparable.

Idealizations that are inherent in the model include the static nature
of the scatterers, the assumption that the scatterers are uncorrelated,
and the single-electron approximation. The effects of static correlations
between the scatterers can be easily estimated from, e.g., Baym's
formula for the electron scattering rate, which can be obtained as a
variational solution of the Boltzmann equation.\cite{Baym} For the
case of static impurities, Baym's formula contains a static impurity
structure factor $S(q)$ under the momentum integral that determines
the scattering rate. For uncorrelated scatterers, $S(q)\equiv 1$. Deviations
of $S(q)$ from unity, and hence static correlations between the
scatterers, are again due to excluded volume effects which we have
estimated above. The dynamics of the Helium gas lead, at temperatures
around $4K$, to corrections of the same order, which can be seen as
follows. A typical value for the scattering time is $\tau\approx 10^{-12}s$.
During this time the thermal velocity of the Helium atoms leads to a
displacement $d$ on the order of $1\AA$. This is comparable with the
scattering length, and the effect on the electron mobility is thus
on the order of $d/\lambda \sim a_s/\lambda$, which is again the
magnitude of the excluded volume effects. Finally, one has to estimate
the effects of the Coulomb interaction, since any real experiment has
to deal with more than one electron. Let us use the parameters of
Schwarz's experiment\cite{Schwarz} for this purpose. A typical value for
the electron current density was
$j\approx 10^{-12}Ccm^{-2}s^{-1}$, and for the drift velocity
$v\approx 10^{4}cm/s$. This
corresponds to an electron density $n_{e} \approx 10^{3}cm^{-3}$.
The Coulomb energy $E_{c}\approx e^{2}/n_{e}^{-1/3}$
is then less than one percent
of the kinetic energy, $k_{B}T$.
If desirable, Coulomb effects can be made even smaller by
decreasing the electron density.

We conclude from the preceding discussion that both the model and
our additional approximations are adequate for a description of
electrons in Helium gas in a parameter range of interest to us,
namely at temperatures and densities such that our expansion,
Eq.\ (\ref{eq:1.3a}), is meaningful. Let us now discuss
the relation of our result to
existing \cite{Schwarz} and possible future experiments. The main
remaining uncertainties in this relation arise from the terms of
$o(\chi^{2})$ and $O(\chi a_{s}/\lambda)$ in Eq.\ (\ref{eq:1.3a}).
The former are undoubtedly nonanalytic, but neither their functional
form nor their magnitude are known.\cite{footnote}
In order to estimate their
importance we therefore have to rely on some assumptions. First,
it is plausible to neglect the nonanalyticity, and to
assume that the coefficients in the $\chi$-expansion are all of
roughly the same magnitude. In the Kubo-Greenwood integration,
Eq.\ (\ref{eq:2.15}), the $\chi^{3}$-term picks up an extra factor of
$\sqrt{\pi}$, and so we estimate $o(\chi^{2})\approx \mu_{3}\chi^{3}$
with $-2\sqrt{\pi}\mu_{2}\alt\mu_{3}\alt 2\sqrt{\pi}\mu_{2}$.
Here we have neglected
the, presumably weak, $\chi$-dependence of $\mu_{3}$, and
have allowed for a
safety margin in the form of an extra factor of $2$. Obviously, the
relative effects of $\mu_3$, and the other unknown terms of higher
order, become smaller with decreasing values of $\chi$. On the other
hand, the excluded volume effects are of
$O(\chi a_{s}/\lambda)$, and become important if $\chi$ becomes
very small. However, as long as
they are small compared to the second order terms which we keep in
Eq.\ (\ref{eq:1.3a}), i.e. as long as $\chi^{2}>\chi a_{s}/\lambda$,
or $\chi>a_{s}/\lambda$, they can be neglected.
This means there is a window of $\chi$-values
for which the excluded volume terms are negligible, but the higher
order in $\chi$-terms are not yet important.

\par
We now define, as a convenient quantity directly comparable with
experiment,
\begin{mathletters}
\label{eqs:3.1}
\begin{equation}
f(\chi) \equiv \left[\mu(T)/\mu_{B} - 1
- \mu_{1}\chi\right]/\chi^{2}\quad.
\label{eq:3.1a}
\end{equation}
Our theoretical prediction for this quantity is,
\begin{equation}
f(\chi) = \mu_{2log} \ln \chi + \mu_{2} \pm \mu_{2} 2\sqrt{\pi} \chi \quad,
\label{eq:3.1b}
\end{equation}
\end{mathletters}%
with $\mu_{2}$ from Eq.\ (\ref{eq:2.16}),
and $\mu_{2log}$ from Eq.\ (\ref{eq:1.3c}).
Let us consider the experimental results obtained by
Schwarz \cite{Schwarz} at Helium temperature. At $T=4.2{\rm K}$, a
He gas density $n=10^{21}cm^{-3}$ corresponds to $\chi=1$, and data
were obtained for $\chi$ as low as $0.08$. In Fig.\ \ref{fig:7} we show
the theoretical prediction, Eqs.\ (\ref{eqs:3.1}), for $0<\chi<0.7$
together with Schwarz's data. The error bars shown assume a total
error of $3\%$ in $\mu/\mu_{B}$ and $4\%$ in $\chi$. To illustrate the
effect of the logarithmic
term the figure also shows what the theoretical
prediction would be if $\sigma_{2log}$ in Eq.\ (\ref{eq:add}) was zero.

We are now in a position to draw the following conclusions. (1) The
existing data are certainly consistent with the existence of the
logarithmic term, but are not accurate enough to be conclusive.
(2) A repetition of this experiment in the region $0.1<\chi<0.2$
with an accuracy improved by at least a factor of $10$ would be
sufficient for a convincing test of the logarithmic term's existence.
This range of $\chi$-values is particularly suitable because excluded
volume effects are negligible.
At larger $\chi$-values the uncertainty
due to the $\chi^{3}$-terms makes the theoretical prediction
meaningless, and at lower $\chi$-values errors in determining $\chi$
translate into very large errors in $f(\chi)$. Also, the excluded
volume effects become noticable at smaller $\chi$. (3) The type
of experiment discussed, i.e. a time-of-flight measurement for electrons
injected into Helium gas, probably constitutes the most favorable
opportunity for an experimental check of the existence or otherwise
of the logarithmic term in the density expansion for transport
coefficients. We hope that the foregoing discussion will stimulate
a new precision experiment on this system.

\acknowledgements

We gratefully acknowledge helpful discussions with P.W.Adams and
J.V.Sengers. One of us (KIW) is grateful for the hospitality
extended to him during a stay at the University of Oregon.
This work was supported by the NSF under
grant numbers DMR-92-17496 and DMR-92-09879.

\appendix
\section{Evaluation of the integrals $J^{++}({\bf \lowercase{q}})$ and
                                         $J^{+-}({\bf \lowercase{q}})$}
\label{app:A}

We have seen in Section II that the diagrams can be
conveniently expressed in terms of the two integrals
$J^{++}(q)$ and $J^{+-}(q)$ given by Eqs.\ (\ref{eqs:2.9}).
In Refs.\ \onlinecite{KirkpatrickDorfman,KirkpatrickBelitz}
approximate representations for these integrals were given, which
are not sufficient for the purposes of the present paper. We therefore
evaluate $J^{++}(q)$ and $J^{+-}(q)$ exactly. The method we are
using actually allows us to calculate two slightly more complicated
functions, viz.
\begin{mathletters}
\label{eqs:A.1}
\begin{equation}
J^{++}(q,\omega) = \int d{\bf k}\
   \frac{1} {\epsilon - {\bf k}^{2} + i\gamma}
   \ \frac{1} {\epsilon + \omega - ({\bf k-q})^2 + i\gamma}\quad,
\label{eq:A.1a}
\end{equation}
and
\begin{equation}
 J^{+-}(q,\omega) = \int d{\bf k}\
  \frac{1} {\epsilon - {\bf k}^2 + i\gamma}
  \ \frac{1} {\epsilon + \omega - ({\bf k} -
  {\bf q})^2 - i\gamma} \quad.
\label{eq:A.1b}
\end{equation}
\end{mathletters}%
$J^{++}(q)$ and $J^{+-}(q)$ are obtained as the values of these
integrals at $\omega =0$.

To proceed we introduce spherical coordinates ($k$, $\theta$,
$\phi$). The integral over the azimuthal angle $\phi$ is
trivial and gives a factor of $2\pi$. The integration over the polar
angle $\theta\in [0,\pi]$ can be written as an integral over
$x = \cos\theta $ extending from -1 to +1. Using the symmetry of the
integrand then allows us to write
\begin{equation}
J^{+\nu}(q,\omega) = 2\pi \int^{1}_{0} dx \int^{\infty}_{0} dk\
 \frac{k^{2}} {\alpha - k^{2}}
 \left[ \frac{1} {\beta_{\nu} - k^{2} - 2 kqx} +
         \frac{1} {\beta_{\nu} - k^{2} + 2 kqx} \right]\quad,
\label{eq:A.2}
\end{equation}
where $\alpha = \epsilon + i\gamma$,
$\beta_{\nu} = \epsilon + \omega + i\nu \gamma - q^{2}$ and $\nu = \pm 1$.
Since the integrand is an even
function of $k$, one can now extend the $k$-integration to the
interval $[-\infty,\infty]$, and evaluate the integral by means of
the residue theorem. The remaining integrals over $x$ are elementary,
and one finds,
\begin{eqnarray}
J^{+\nu}(q,\omega) =
 \nu\ \frac{i\pi^{2}} {2q}\
 \left[\ln\left(\frac{\sqrt{q^{2} +\beta_{\nu}} + q}
                  {\sqrt{q^{2} + \beta_{\nu}} - q}\right)\right.\nonumber\\
  - \left. \ln\left(
  \frac{2\nu\sqrt{\alpha} \beta_{\nu} + q(\alpha - \beta_{\nu}) +
           (\alpha + \beta_{\nu}) \sqrt{q^{2} + \beta_{\nu}}}
       {2\nu\sqrt{\alpha} \beta_{\nu} - q(\alpha - \beta_{\nu}) +
           (\alpha +\beta_{\nu})\sqrt{q^{2} + \beta_{\nu}}}\right)\right]\quad.
\label{eq:A.3}
\end{eqnarray}
After some simplifications we obtain,
\begin{mathletters}
\label{eqs:A.4}
\begin{equation}
J^{++}(q,\omega) = -\frac{i\pi^{2}} {q}\
  \ln\left( \frac{\sqrt{\epsilon +
                \omega + i\gamma} + \sqrt{\epsilon + i\gamma} - q}
           {\sqrt{\epsilon +
            \omega + i\gamma} + \sqrt{\epsilon + i\gamma} + q}\right)\quad,
\label{eq:A.4a}
\end{equation}
and,
\begin{equation}
J^{+-}(q,\omega) = \frac{i\pi^{2}} {q}\
  \ln\left(\frac{\sqrt{\epsilon + \omega - i\gamma} - \sqrt{\epsilon +
                   i\gamma} - q}
           {\sqrt{\epsilon + \omega - i\gamma} - \sqrt{\epsilon +
              i\gamma} + q}\right)\quad.
\label{eq:A.4b}
\end{equation}
\end{mathletters}%
In the limit $\omega\rightarrow 0$, $J^{+-}$ can be further simplified,
and we finally obtain,
\begin{mathletters}
\label{eqs:A.5}
\begin{equation}
J^{++}(q) = \frac{i\pi^{2}} {q} \ln\left(
   \frac{2\sqrt{\epsilon + i\gamma} + q}
         {2\sqrt{\epsilon + i\gamma} - q}\right)\quad,
\label{eq:A.5a}
\end{equation}
and
\begin{equation}
J^{+-}(q) = \frac{2\pi^{2}} {q} \arctan\left(
    \frac{q} {2{\rm Im}\sqrt{\epsilon + i\gamma}}\right)\quad.
\label{eq:A.5b}
\end{equation}
\end{mathletters}%
For small $\gamma$ these expressions agree with those obtained
previously.\cite{KirkpatrickBelitz}

For explicit perturbative calculations the following formal
small-$\gamma$ expansions are useful,
\begin{mathletters}
\label{eqs:A.6}
\begin{equation}
J^{++}(q) = \frac {\pi^3} {q} \Theta(q-2\sqrt{\epsilon}) - i \frac {\pi^2}
{q} \ln \left\vert \frac { q-2\sqrt{\epsilon}} {q+2\sqrt{\epsilon}}
\right\vert
+ {\gamma \over \epsilon} \left( \frac {2\pi^2 \sqrt{\epsilon}}
{4\epsilon -q^2} -i \pi^3 {\sqrt{\epsilon} \over q }
\delta(q-2\sqrt{\epsilon})\right) + O(\gamma^2)\quad,
\label{eq:A.6a}
\end{equation}
\begin{equation}
J^{+-}(q) = {\pi^3 \over q } - \frac {\gamma} {\sqrt{\epsilon}}{ 2\pi^2
\over q^2} + O(\gamma^2)\quad.
\label{eq:A.6b}
\end{equation}
\end{mathletters}%

\section{Evaluation of diagram 2(\lowercase{a})}
\label{app:B}

To calculate the contribution of diagram 2(a) to
the zero-temperature conductivity we use the expressions for
$J^{++}(q)$ and $J^{+-}(q)$ from Appendix \ref{app:A} in
Eq.\ (\ref{eq:2.7}), which we rewrite as,
\begin{eqnarray}
\sigma_{(2a)} = \sigma_{B} \left(\frac{\gamma} {2\epsilon}\right)
   \frac{1} {2\pi^5\epsilon^{3/2}}\biggl[
  {\rm Re}\ (-i\gamma)\ \int_0^Q dq\ q^2\ \left(J^{++}(q)\right)^2
                                                               \nonumber\\
 +{\rm Re}\ (i\gamma)\ \int_0^Q dq\ q^2\ J^{++}(q)\,J^{+-}(q)
 -\epsilon\ {\rm Re}\ \int_0^Q dq\ q^2\ J^{++}(q)\,J^{+-}(q)
                                                               \nonumber\\
 -\epsilon\gamma\ {\rm Re}\ \int_0^Q dq\ q^2\ J^{++}(q)\,
                                               {d\over d\gamma}\,J^{++}(q)
 +\epsilon\ {\rm Re}\ \int_0^Q dq\ q^2\ \left(J^{++}(q)\right)^2
 \biggr]\ +\ O(\gamma^2)\quad.
\label{eq:B.1}
\end{eqnarray}
Because there is a factor $\gamma/2\epsilon$ multiplying the
integrals it is sufficient to evaluate the latter to linear order
in $\gamma$. Of the five integrals the first two can be straightforwardly
evaluated by using the expansion given in Eq.\ (\ref{eqs:A.6}). The same
is true for the third integral, although the expansion procedure leads
to an ill-defined integral $\int_0^2 dx/(x-1)$, which has to be
interpreted in a principal values sense (so it is zero). Alternatively,
one can use the exact expressions for $J^{++}$ and $J^{+-}$ from
Eqs.\ (\ref{eqs:A.5}), write the $\arctan$ as an auxiliary integral,
$\arctan x = x\int_0^1 dy/(1+y^2x^2)$, and use complex analysis. The
result is the same, viz.
\begin{equation}
{\rm Re}\ \int_0^Q dq\ q^2\ J^{++}(q)\,J^{+-}(q) =
 \pi^6(Q-2{\rm Re}\sqrt{\epsilon + i\gamma})
 -\left(\gamma\,4\pi^5/\sqrt{\epsilon}\right)\ln (Q/2\sqrt{\epsilon})
 + O(\gamma^2)\quad,
\label{eq:B.2}
\end{equation}
where we have dropped terms that vanish for $Q\rightarrow\infty$.
The fourth term requires the evaluation of an integral,
\begin{mathletters}
\label{eqs:B.3}
\begin{equation}
J\equiv\lim_{\delta\rightarrow 0}\ {\rm Im}\ \int_{-Q}^Q dq\ q\
 \ln\left(\frac{2\sqrt{\epsilon+i\gamma}+q}
                                        {2\sqrt{\epsilon+i\gamma}-q}\right)\
 \frac{1}{q+2\sqrt{\epsilon+i\gamma}+\delta}\
 \frac{1}{q-2\sqrt{\epsilon+i\gamma}-\delta}\quad,
\label{eq:B.3a}
\end{equation}
where we have made use of the symmetry of the integrand, and have
shifted the poles off the branch cuts of the logarithm.
Standard complex analysis techniques yield, in the limit of large $Q$,
\begin{equation}
J = -2\pi\left[\ln(Q/2\sqrt{\epsilon}) - \ln 2\right] + O(\gamma^2)\quad.
\label{eq:B.3b}
\end{equation}
\end{mathletters}%
Finally, the fifth integral can be related to the fourth one by Taylor
expanding in $\gamma$. Adding all contributions one obtains entry
2(a) in Table \ref{tab:1}.

\begin{figure}
\caption{The simple bubble (a), and two diagrams containing the
 'triangulated' Green's function (b), (c). Together with diagrams
 (b) and (c), their complex conjugates (c.c.) also contribute.}
\label{fig:1}
\end{figure}

\begin{figure}
\caption{Infinite resummations that are derivatives of diagrams 2(d)-2(h).}
\label{fig:3}
\end{figure}

\begin{figure}
\caption{The Green's function to second order in the impurity density}
\label{fig:6}
\end{figure}

\begin{figure}
\caption{The reduced mobility $f$, as defined in
 Eq.\ (\protect\ref{eq:3.1a}),
 vs. the density parameter $\chi=\lambda/\pi l$. The theoretical
 prediction is for $f$ to lie between the two solid lines. The
 experimental data are from Fig. 9 of Ref.
 \protect\onlinecite{Schwarz} with
 error bars estimated as described in the text. The broken lines
 show what the theoretical prediction would be in the absence of
 the logarithmic term in the density expansion.}
\label{fig:7}
\end{figure}

\begin{table}
\caption{Values of the diagrams as shown in
 Figs.\ \protect\ref{fig:1}-\protect\ref{fig:5}.
 Only the contribution to the coefficient $\sigma_2$ in
 Eq.\ (\protect\ref{eq:add}) is given, for contributions to
 lower orders see
 Refs.\ \protect\onlinecite{KirkpatrickDorfman,KirkpatrickBelitz}.
 See the text for further explanation}
\begin{center}
\begin{tabular}{cccc}
diagram & $\sigma_2$ & diagram & $\sigma_2$ \\ \hline
 1\ (a) & $3/2$ &  4\ (f) & $- I_{1} + 2I_{2}$ \\
 1\ (b) & $-7/2$ &  4\ (g) & $4 \ln 2 - 2$ \\
 1\ (c) & $1$ &  4\ (h) & $2 - 4 \ln 2$ \\
 2\ (a) & $6 \ln Q/\sqrt{\epsilon}+4-1 2\ln 2$ &  4\ (i) & $I_{1}/2 - I_{2}$ \\
 2\ (b) & $8$ &  5\ (a)+(b) & $0$ \\
 2\ (c) & $- 2 - 4 \ln 2$  &  5\ (c)+(d) & $0$ \\
 2\ (d) & $\pi^{2}/2 + 3I_{1}/2$ &  5\ (e) & $\pi^{2}/9$ \\
 2\ (e)+(f)+(g)+(h) & $- 1 - \pi^{2} - 2 \ln 2$ &  5\ (f) & $0$ \\
 3\ (a)+(b)+(c) & $\pi^{2}/4 - 2 \ln 2$ &  5\ (g) & $2\pi^{2}/3$ \\
 3\ (d) & $-I_{3}$ &  5\ (h)+(i) & $\pi^{2}$ \\
 4\ (a) & $- I_{1}/2$ &  5\ (j)+(k) & $0$ \\
 4\ (b) & $- I_{1}/2$ &  5\ (l)+(m) & $0$ \\
 4\ (c) & $- I_{1}$ &  5\ (n)+(o) & $0$ \\
 4\ (d) & $-2 \ln(Q/\sqrt{\epsilon}) + 4 \ln 2$ &  5\ (p) & $0$ \\
 4\ (e) & $2\ln 2 - 1$ & \, & \, \\
\end{tabular}
\end{center}
\label{tab:1}
\end{table}


\begin{references}
\bibitem{footnote} Our calculations suggest that the leading nonanalyticity
 at third order in $\chi$ is of the form $\chi^3\,(\ln\chi)^2$.
\end{references}
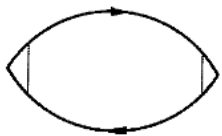
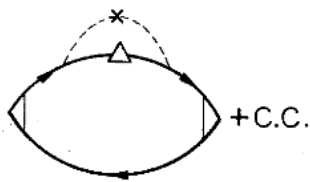
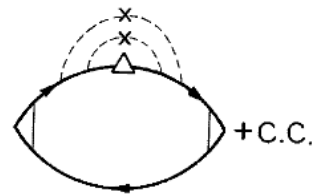
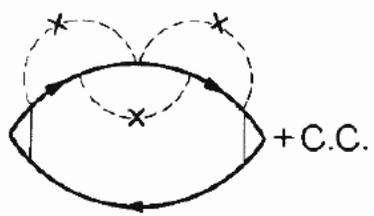
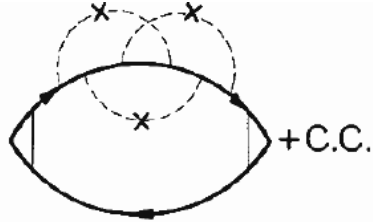
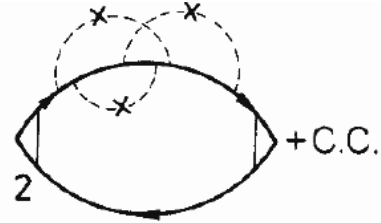
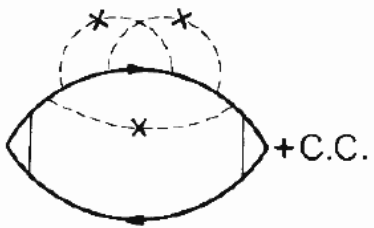
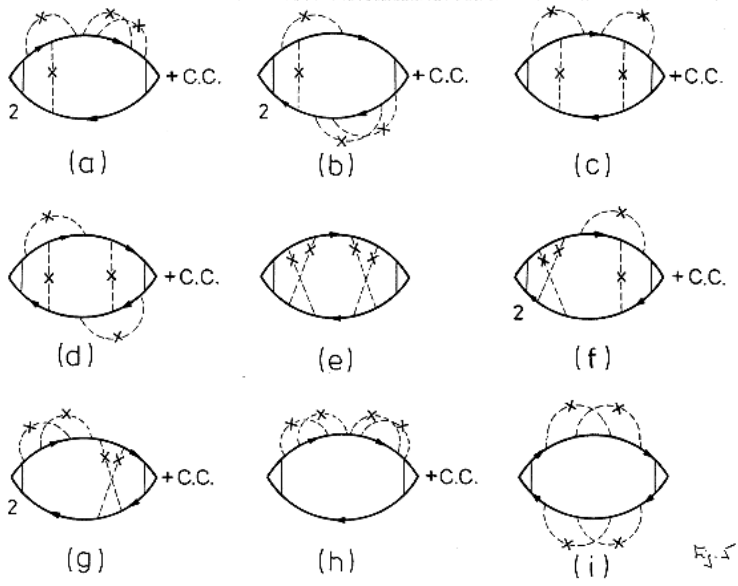
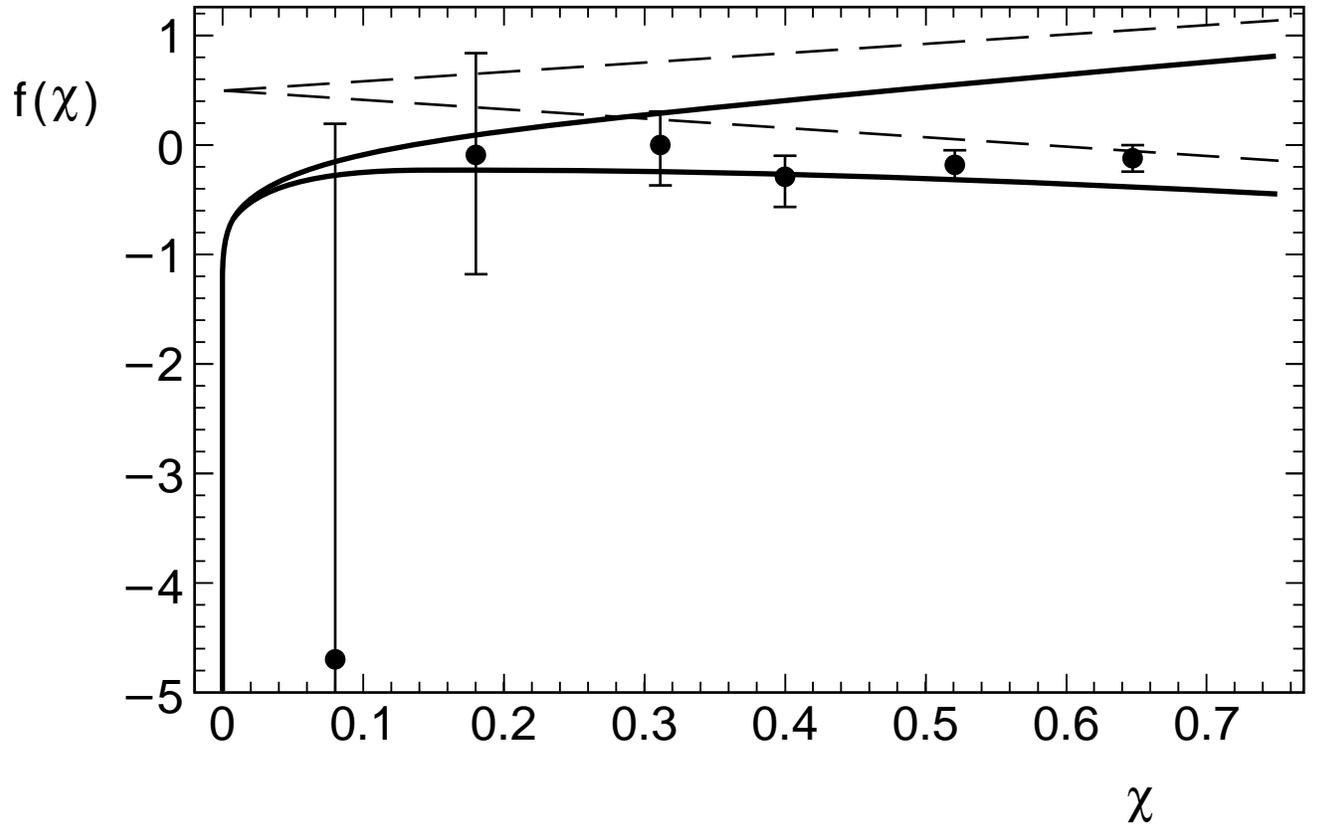
\end{document}